\documentclass[article,prd]{revtex4}

\usepackage{amsbsy}
\usepackage{amssymb}
\usepackage{amsmath}
\usepackage{graphicx}

\def\x{{\mathrm{x}}}
\def\y{{\mathrm{y}}}
\def\n{{\rm n}}
\def\p{{\rm p}}

\def\s{\mathrm{s}}
\def\f{{\rm f}}
\def\I{{\rm c}}

\def\b{{\rm b}}
\def\S{{\mathrm S}}
\def\v{{\mathrm v}}
\def\N{{\mathrm N}}

\def\bep{\bar{\varepsilon}}
\def\bB{\bar{\mathcal{B}}}
\def\bBp{\bar{\mathcal{B}}'}

\newcommand{\be}{\begin{equation}}
\newcommand{\ee}{\end{equation}}

\begin{document}

\title{A superfluid perspective on neutron star dynamics}

\author{N. Andersson}

\affiliation{Mathematical Sciences and STAG Research Centre, University of Southampton,
Southampton SO17 1BJ, United Kingdom}

\begin{abstract}
As mature neutron stars are cold (on the relevant temperature scale), one has to carefully consider the state of matter in their interior. The outer kilometer or so is expected to freeze to form an elastic crust of increasingly neutron-rich nuclei, coexisting with a superfluid neutron component, while the star's fluid core contains a mixed superfluid/superconductor. The dynamics of the star depend heavily on the parameters associated with the different phases. The presence of superfluidity brings new degrees of freedom---in essence we are dealing with a complex multi-fluid system---and additional features: Bulk rotation is supported by a dense array of quantised vortices, which introduce dissipation via mutual friction, and the motion of the superfluid is affected by the so-called entrainment effect. This brief survey  provides an introduction to---along with a commentary on our current understanding of--- these dynamical aspects, paying particular attention to the role of  entrainment, and outlines the impact of superfluidity on neutron-star seismology. 
\end{abstract}

\maketitle

\section{Neutron star superfluidity}

During its first moments of existence, just after the supernova core collapse,  a newly born neutron star 
can be thought of as a hot, rotating, ``ball'' of superdense nuclear material. Thermal aspects play a key role during the early stages of evolution, and it is essentially the loss of the associated pressure support (through the emission of neutrinos) that leads to the newly born object shrinking from a radius of about 20~km to the typical size of  a ``mature'' neutron star; likely just over 10~km. Once the temperature drops below about $10^{10}$~K (or 1~MeV for anyone of a nuclear physics persuasion)---after the first 20-100~s \cite{Lattimer}---the thermal contribution to the pressure may be ignored and we can meaningfully consider the object as a---now gradually evolving---neutron star. From the nuclear physics point of view, the object is cold. In fact, it so cold that we have to consider the precise state of matter. 

\begin{figure}
\begin{center}
\includegraphics[width=0.7\columnwidth]{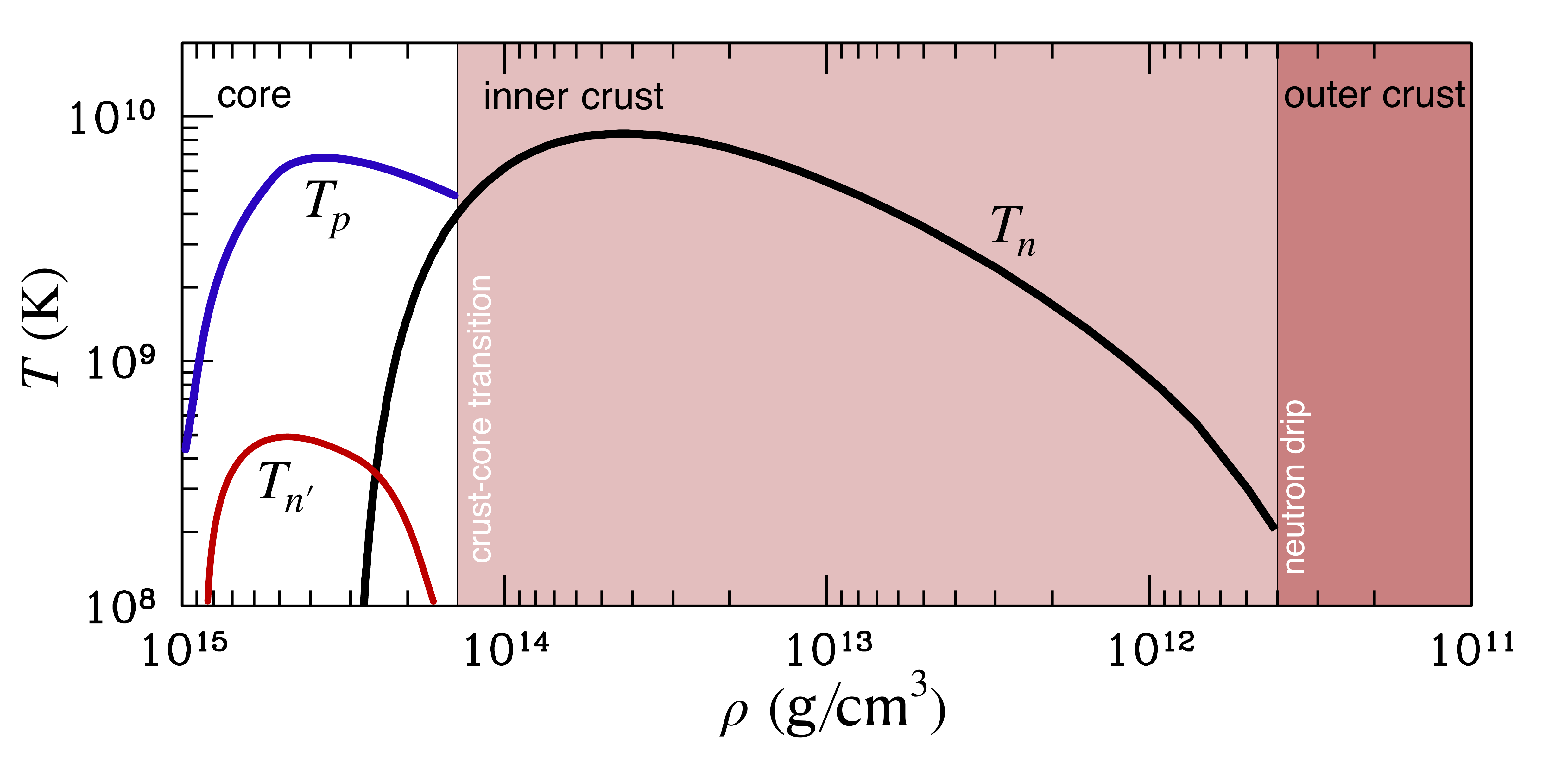}
\end{center}
\caption{A sketch of the critical temperatures for superfluidity/superconductivity as a function of density. The indicated critical temperatures $T_n$ (black), $T_{n'}$
(red), and $T_p$ (blue) represent neutron singlet,
neutron triplet, and proton singlet pairing, respectively. The least ``secure'' of these critical curves is that for neutron triplet pairing. The edges of the coloured regions indicate the  boundaries between the
core and the inner crust (at just above half the  nuclear saturation density) as well as the inner and outer crusts (at neutron drip;
$\rho\approx 4\times 10^{11}$~g\,cm$^{-3}$). The main message is that superfluid aspects come into play already at a temperature close to $10^{10}$~K (about 1~MeV), very early in a neutron star's life. ({The results provide a phenomenological representation of actual gap calculations within the BCS approximation, see \cite{supercool} for details.)}} 
\label{pair}
\end{figure}

Laboratory experiments tell us that two things may happen when we cool a fluid. It may freeze---as in the familiar winter-time example of water forming ice---or it may become superfluid---as in the case of low-temperature laboratory experiments on Helium. The latter outcome is less common, as it requires quantum fluctuations to prevent the formation of a regular particle lattice. However, neutron stars---obviously, not hands-on laboratories!---are expected to manifest both phases. The outer kilometer or so forms the star's crust, a lattice composed of increasingly neutron-rich
nuclei \cite{chamrev}. The outer core of the star is expected to contain a mixture of superfluid neutrons---forming a condensate due to an analogue of Cooper pairing below a density-dependent critical temperature, see figure~\ref{pair}---alongside a charge neutral conglomerate of protons and electrons (with muons also coming into play as the density increases). This may already seem a fairly complicated system, but we need (at least) two further features. First, beyond a density of about $4\times10^{11}$~g/cm$^3$, neutrons start to drip out of crust nuclei, leading to a superfluid coexisting with the crust lattice. Second, the protons in the outer core are likely to form a superconductor. If we proceed to the very high densities of the deep neutron star core, we may have to consider the impact of hyperons and/or deconfined quarks \cite{glend,asrev}. These are also expected to become superfluid, but we will not consider the additional issues this gives rise to here. 

The notion of neutron star superfluidity is firmly established theoretically and---as our observations become more precise---the observational support for the idea is getting stronger. Presently,  the observations relate to three distinct problems: (i) the transient cooling following phases on matter accretion onto the neutron star surface, for which the crust superfluid regulates the thermal relaxation timescale \cite{crustcool}, (ii) the apparent  real-time cooling of the young neutron star in the Cassiopeia A supernova remnant, which (if confirmed) requires the recent onset of core superfludity \cite{casa1,casa2}, and (iii) the enigmatic spin glitches seen in many young radio pulsars \cite{baym69,ai75,espinoza11}. As the present focus is on dynamical aspects, we will only consider the last of these aspects here.

Even though the problem of neutron star superfluidity has been under scrutiny since the late 1960s (following the original suggestion of Migdal \cite{mig}), many relevant aspects remain unresolved (see \cite{rev1,rev2,rev3} for recent reviews). These range from nuclear physics issues, like the pairing \cite{pair1,velgaps} and the critical temperature (with the neutron triplet pairing gap---roughly the binding energy of a Cooper pair---still uncertain) and the interaction between quantised neutron vortices and crust nuclei leading to vortex pinning \cite{pin1,pin2,pin3,pin4,pin5,pin6,pin7,pin8}, to large-scale dynamics and the impact on astrophysical observations, with the mechanism that triggers pulsar glitches remaining not well understood. This brief survey is intended to serve an introduction to dynamical aspects that come into play when we consider a superfluid neutron star. This involves additional degrees of freedom---we are dealing with a complex multi-fluid system---and new features like the dense array of quantised vortices that allows a superfluid to mimic bulk rotation and the associated mutual friction dissipation. In addition, new parameters  need to be understood. Of particular importance in this respect is the so-called entrainment effect \cite{andreev75:_three_velocity_hydro,als,entr1,entr2,entr3}, the relevance of which will be a recurring motif throughout this discussion.
 
\section{The essence of the two-fluid model}

The effort to model the dynamics of superfluid neutron stars draws inspiration from laboratory systems. Historically,  the thinking was based on Landau's original two-fluid model for He$^4$ \cite{landau59:_fluid_mech} (see also \cite{khalatnikov,wilks,putterman}),  with aspects relating to experiments on cold atom gases becoming increasingly relevant more recently. There are many conceptual similarities,  see \cite{vanessa} for a detailed discussion, but the fineprint details are quite different. 
In the case of neutron stars, both neutrons and protons are expected
to form Cooper-type pairs, which then form condensates that can be described by fluid equations. A key distinction from  Helium is that the two components no longer represent the 
``superfluid'' and ``normal'' parts of a single particle species.  Instead, the 
two degrees of freedom describe (in the simplest case, which we focus on here) the neutrons and a conglomerate of all 
charged components (which are expected to be electromagnetically coupled 
on a short timescale). 

The dynamical equations for the system have to faithfully represent the degrees of freedom and the interaction between them. As we will see, there are different ways to do this. In essence, the models involve an element of choice (or perhaps, taste). The modern approaches to the problem reflect this choice, but they are all related in a fairly straightforward manner so one can translate the results from one paradigm to another \cite{LivRev} (we  provide specific examples of this  later). The ability to complete such translations is, of course, crucial. It should not  be the case that the physics depend on our mathematical perspective. 

The traditional approach was adopted in the seminal effort of Mendell 
\cite{mendell1,mendell2}, who
extended the non-dissipative 
zero-temperature equations to account for the main dissipation
mechanisms for neutron star cores. His work remains a guiding beacon for research in this area. The final set of  equations include the mutual 
friction coupling due to electrons scattering off of rotational vortices 
in the condensates (see also \cite{als,mutualf}). As the mathematical framework was applied to the problem of  unstable modes of oscillation in a 
spinning neutron star it became clear that the 
mutual friction is crucial for an understanding of neutron star dynamics. The mechanism may, in fact,  
suppress (some) rotational instabilities entirely 
\cite{lindblom95}. Mendell's work also demonstrates how the model is complicated by the  so-called entrainment 
effect \cite{andreev75:_three_velocity_hydro,als,entr1,entr2,entr3}, which 
accounts for the fact that the flow of one fluid may impart 
momentum in another. As the entrainment features prominently in discussions of superfluid neutron star dynamics, we will pay particular attention to its origin and interpretation in the following. In fact, this will be the main thread of the discussion.

\subsection{The equations of motion}

In order to provide an intuitive impression of the equations that govern superfluid neutron stars, we will focus on the Newtonian equations. It is well known that realistic neutron stars require general relativity for a detailed description, but the main features of the fluid dynamics do not change conceptually---the keen reader may confirm this by consulting the material in \cite{LivRev}---so we can introduce the ideas within the somewhat simpler Newtonian setting. The starting point for our description is different from that of Mendell \cite{mendell1,mendell2} in that we make a clear distinction between transport velocities and momenta for the different components of the system. One reason for doing this is that the relations we write down then follow as the low-velocity/weak-field limit of the corresponding relativistic model \cite{LivRev}. 

There are, in fact, three different ways to obtain these equations. Perhaps the most intuitive is to develop the relativistic theory and then work out the Newtonian limit \cite{LivRev}. Working entirely within Newtonian physics, it is possible to develop a variational framework (involving time-shifts to mimic the Lagrangian variations from the relativistic model) for deriving the equations \cite{prix}. There is also a hybrid approach, centered on the Milne spacetime structure and the Cartan connection (involving a degenerate metric) 
\cite{carter04:_cov1,carter03:_cov2,carter04:_cov3}. A third alternative is provided in \cite{twof}. One may also consider a general approach for adding dissipation to the system, drawing on Onsager's celebrated symmetry principle \cite{monster}.

Making use of a chemical label $\x,\y$ (which does not follow the Einstein summation convention), the simplest relevant superfluid neutron star model involves two components, distingushing the superfluid neutrons, 
with mass density $\rho_\n$ and velocity $v^i_\n$, from the ``protons'', a 
conglomerate of all charged components, represented by
$\rho_\p$ and $v^i_\p$. 
In absence of dissipation, each component is represented by a continuity equation (for mass conservation)
\begin{equation}
\partial_t \rho_\x + \nabla_i \left(\rho_\mathrm{x} v_\mathrm{x}^i\right) = 0 
\label{continue}
\end{equation}
where $\mathrm{x}=\n$ or $\p$.
Here, and in the following, we express the equations in terms of components in a coordinate basis, which means that $\nabla_i$ should be viewed as the corresponding covariant derivative. Momentum conservation leads to
\begin{equation}\label{euler}
\left(\partial_t + {v}^{j}_{\x}\nabla_{j} \right)   
\left( {v}_{i}^{\x} + \varepsilon_{\x} w_i^{\y\x} \right) + \nabla_{i} (\Phi + 
\tilde{\mu}_{\x}) 
+ \varepsilon_{\x}  w_j^{\y\x} \nabla_{i} v^{j}_{\x} = 0 \ .
\end{equation}
where  $\y\neq\x$ and the relative velocity between the components---an essential part of the story---is defined as
\begin{equation}
w_i^{\y\x} =  {v}_{i}^{\y} - {v}_{i}^{\x} \ . 
\end{equation}
In this description, the matter equation of state is represented by a Lagrangian (see \cite{monster} for details)
\begin{equation}
\mathcal L = \sum_\x {m_\x \over 2} g_{ij} n_\x^i n_\x^j - \mathcal E(n_\x, n_\x^i)
\end{equation}
where $n_\x^i = n_\x v_\x^i$ and the internal energy $\mathcal E$ notably depends not only on the matter densities, it must also involve the relative velocity. Assuming $\mathcal E=\mathcal E(n_\n, n_\p, w^2_{\n\p})$, 
the chemical potential for each particle species is given by 
\begin{equation}
\mu_\x = m_\x \tilde \mu_\x = \left( { \partial \mathcal E \over \partial n_\x}\right)_{n_\y, w_{\n\p}^2} \ ,
\end{equation}
with $m_\x$ the mass of each particle. We also have the entrainment coefficients
\begin{equation}
\varepsilon_\x = {2 \alpha \over \rho_\x}  \qquad  \mbox{ where } \qquad \alpha = \left({ \partial \mathcal E \over \partial w_{\n\p}^2} \right)_{n_\x}\ .
\end{equation}
Finally,
$\Phi$ represents
the gravitational potential (as we want to model stars!).

In order to appreciate the impact of the entrainment, it is useful to note that the fluid momentum is given by
\begin{equation}
p_i^\x = \left( {\partial \mathcal L \over \partial n_\x^i} \right)_{n_\x} =   m_\x \left( {v}_{i}^{\x} + \varepsilon_{\x} w_i^{\y\x} \right)  \ , 
\label{mom_def}\end{equation}
This illustrates the key point: 
 The momentum of  
each fluid need not be parallel with the fluid's transport velocity. The expression for the momentum \eqref{mom_def}  suggests an intuitive way to think of this. We may define the effective proton mass  in a frame moving with the neutrons (setting $v_\n^i=0$):
\begin{equation}
m_\p (1-\varepsilon_\p) v_\p^i \equiv m_\p^* v_\p^i
\end{equation}
with a similar relation for the neutrons defining $m_\n^*$. We now have
 \begin{equation}
2 \alpha = \rho_\p \varepsilon_\p =  n_\p (m_\p - m_\p^*) \equiv n_\p \delta m_\p^* \ .
\end{equation}
Moreover, it i easy to see that the effective masses must be related by
\begin{equation}
m_\n-m_\n^* = {n_\p \over n_\n } (m_\p - m_\p^*)
\end{equation}

The notion of a dynamical effective mass is important as it encodes a key effect expected from nuclear physics. 
In a neutron star core, the strong force
endowes each neutron with a virtual cloud of protons (and vice versa). 
This affects the effective mass of the particle and, when it moves, alters 
the momentum. The mass of each neutron appears different from what it would be in isolation.
It is important to stress that the dynamical effective masses  are distinct from the static  (Landau) effective masses in nuclear physics calculations.  Having said that, the two sets are close for systems with small proton fractions (see the discussion in  \cite{chamrev}). The upshot is that, when we model neutron star cores we can to good precision ignore the somewhat technical point that explains the difference. This was, indeed, done in the analysis that led to the results shown in figure~\ref{entrain}.

An alternative description \cite{carter04:_cov1,carter03:_cov2,carter04:_cov3} introduces a symmetric ``mobility matrix'' $\mathcal K^{\x\y}$ such that
\begin{equation}
p_i^\x = \mathcal K^{\x\x} n^\x_i  + \mathcal K^{\x\y} n^\y_i
\end{equation}
It is easy to see that we then identify
\begin{equation}
m_\p^* = n_\p \mathcal K^{\p\p}
\end{equation}
and
\begin{equation}
\alpha = {1\over 2} n_\n n_\p \mathcal K^{\n\p}
\end{equation}
This picture highlights the interpretation of the entrainment as a measure of how easy it is to induce a relative flux between the two fluids; in essence, how mobile the superfluid neutrons are relative to the protons (and vice versa). At the end of the day, the two descriptions are (obviously) equivalent and point to a model that requires three equation of state parameters: the two number densities $n_\n$ and $n_\p$ along with a measure of the entrainment, like $\alpha$, $m_\p^*$ or $\mathcal K^{\n\p}$.

\subsection{The crust and the chemical gauge}

The equations we have considered so far apply in the fluid outer core of a neutron star (provided we ignore electromagnetism). In this region,  the
distinction between the two components is  clear. We count, on the one hand, all the neutrons and, on the other hand, everything else. The inner crust region requires additional consideration---the problem is both similar and different. The similarity is obvious at a glance: 
At densities beyond neutron drip, some neutrons
remain bound in nuclei but there is also a ``gas'' of free neutrons.
The assignation of neutrons to each component follows from the
equation of state, once the nature of the ions in the lattice is
established \cite{chamrev}. We are dealing with a two-component system which may exhibit a relative flow. However, if we take a closer look, we see that there are complicating aspects. In particular, it
is not clear to what extent the neutrons that are  ``confined'' to the crust nuclei are able to
move \cite{cacha}. This depends on how strongly bound they are, to what extent
they can tunnel through the relevant interaction potentials
and so on. 

An intuitive discussion of this problem introduces the notion of 
chemical ``gauge'' \cite{cacha}. The chemical gauge involves the
interpretation of the different quantities (number densities, etc), while the two-fluid model remains unchanged at the formal level. The idea is somewhat abstract, so let us focus on the situation in the inner crust. 
To make a
distinction from the outer core setting, we  refer to the two
components in the crust as ``free'' neutrons, with density $n_\f$,
and ``confined'' baryons, represented by $n_\I$ \cite{cacha,casa}. We also need to consider the number density of baryons associated with crust nuclei, $n_\N$, (representing all protons as well as the confined neutrons, making up the ions in the lattice). The point is that this set-up is distinct from one where we count all neutrons, $n_\n$. To make things clear(er) we need to consider the physical meaning of $n_\f$ and $n_\N$. We also need to explain why we make a distinction between the core and the crust. The answer is easy. In the crust, we have to consider the additional restoring force due to the elastic lattice. The elastic restoring force involves the nuclei, and hence the confined neutrons. The component that is free to move, is  represented by $n_\f$. Of course, the entrainment blurs the distinction.
Still, the option to consider a  two-fluid model based on $n_\n$ and $n_\p$ also in the crust is not very attractive as one would then have to split the elastic contribution between the two momentum equations and this would be (even more) confusing.

\begin{figure}
\begin{center}
\includegraphics[width=0.7\columnwidth]{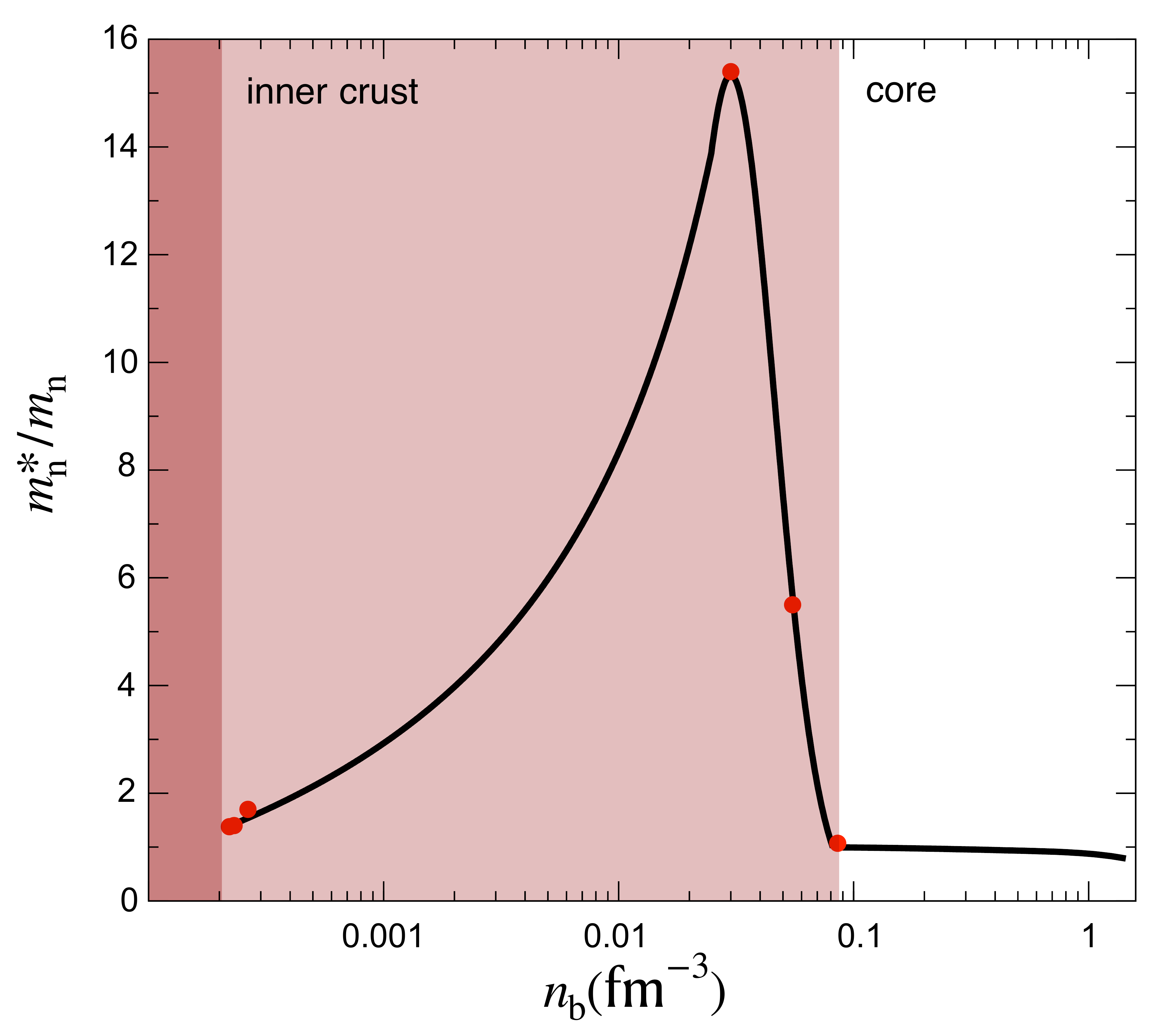}
\end{center}
\caption{An indication of the dependence of the effective neutron mass---encoding the superfluid entrainment---on
  the baryon number density, $n_\b$. {In the crust, the  curve has been fitted to the data points from Chamel and
  collaborators~\citep{crent0,crent5,crent6} (shown as filled circles)}.  The results in the core are obtained by assuming that the dynamical effective mass is equal to the Landau effective mass from nuclear physics (which should be a good approximation as the proton fraction is small).} 
\label{entrain}
\end{figure}

In essence, the chemical gauge relates to the neutrons that are considered ``free''.
The issue is subtle since, in a dynamical situation,  the neutrons that are associated with the nuclei may be able to tunnel through
the relevant interaction potential. This, in turn, makes concepts like the atomic number less precise. In general, one may introduce a new
basis such that
\begin{equation}
n_\f^i = n_\n^i + (1-a_\N) n_\p^i
\end{equation}
where $a_\N$ (which we will take to be constant in the following; a good approximation at the level of the individual fluid elements \cite{cacha})
accounts for the fact that some of the neutrons move with the
(crust) protons. We also have
\begin{equation}
n_\N^i = a_\N n_\p^i
\end{equation}
Given these relations, it is easy to show that the neutron momentum is independent of the chosen chemical gauge, represented by $a_\N$ \citep{cacha}.
This follows immediately from the definition of the momentum \eqref{mom_def}. We then have $p^\n_i = p^\f_i$ and $\mu_\n = \mu_\f$.
These are important identifications, but it is  clear that we must in general have $v_\f^i \neq v_\n^i$ and $\varepsilon_\f \neq \varepsilon_\n$.
In essence, we have to execute some care when we discuss the entrainment and the dynamics of the neutron star crust.

The discussion of chemical gauge is reflected by different models for crust dynamics discussed in the literature, see for example \cite{crust1,crust2,crust3,crust4,crust5,crust6,crust7}. At first sight, various descriptions may appear to differ but a close inspection shows that they have the same physics content. One of the key recent developments is the realisation that the superfluid neutrons in the crust may not be particularly mobile. Evidence suggests that the superfluid neutron component is strongly coupled to the crust lattice via Bragg scattering \cite{crent0,crent5,crent6,crent1}. As a result, the effective neutron mass may be as much as an order of magnitude larger than the bare mass at some points in the crust, see the results shown in figure~\ref{entrain}. As we will argue later, this result---which continues to be debated \cite{crent2,crent3,crent4,crent7}---could have significant impact on the dynamics of the system.

\subsection{Thermal excitations}

\begin{figure}
\begin{center}
\includegraphics[width=0.7\columnwidth]{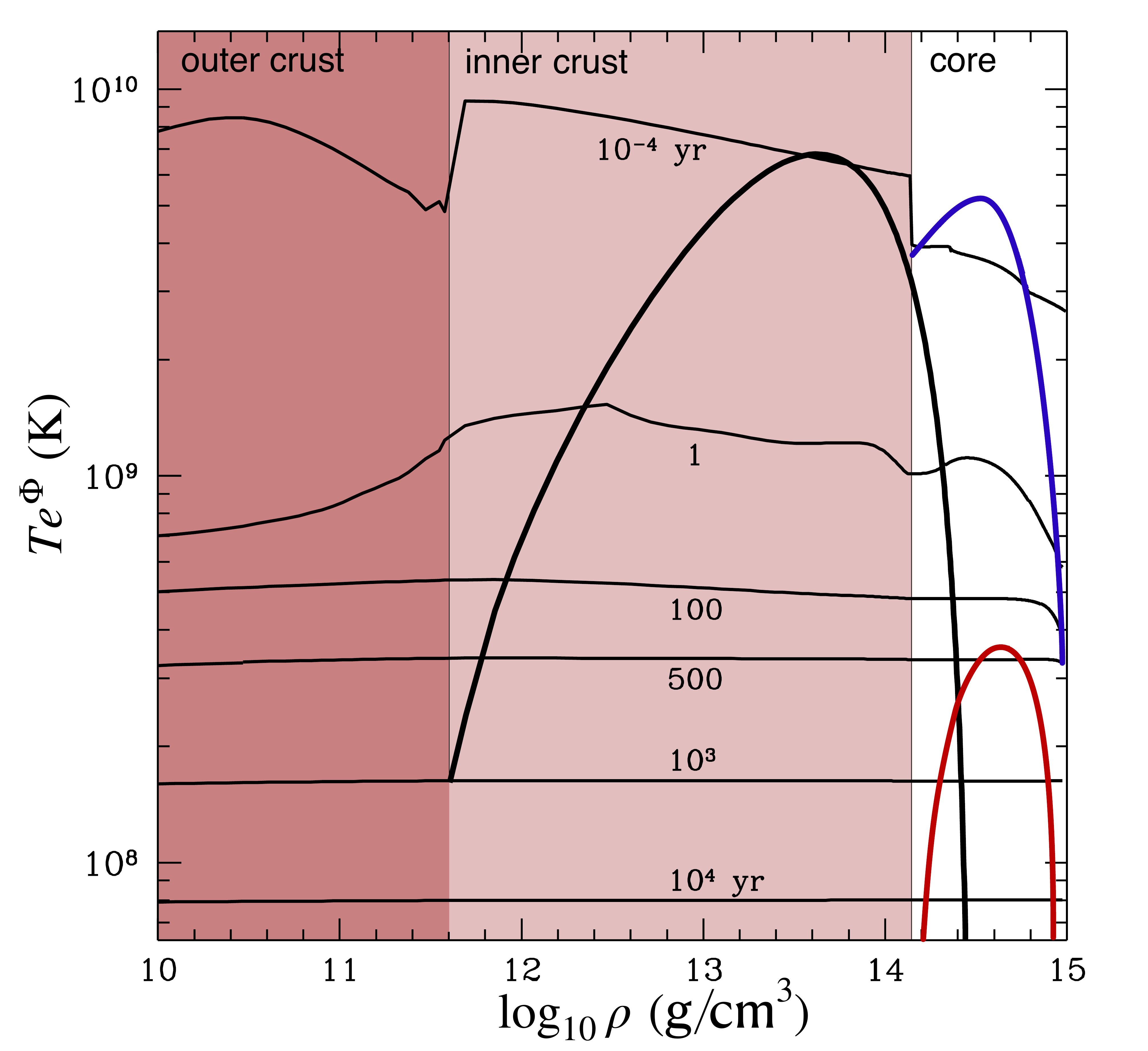}
\end{center}
\caption{Temperature profiles for a neutron star cooling model with superfluidity and no additional heating
(thin black curves).
Six (redshifted) temperature profiles are shown, representing  ages $10^{-4}$ (top), 1, 100, 500, $10^3$, and
$10^4\mbox{ yr}$ (bottom), respectively.
The critical temperatures for superfluidity/superconductivity, as well as the different regions in the star, are also shown. The stellar model is the same as is figure~\ref{pair} (reproducing the results from \cite{supercool}.)} 
\label{cool}
\end{figure}

Moving on, let us consider another complicating aspect. So far we have taken the view that neutron stars are cold and that our mission is to model superfluids well below the critical transition temperature. Locally, this makes sense, but  we have to be a bit careful when we consider a global model. This is apparent from results for neutron star cooling, see the example provided in figure~\ref{cool} (taken from \cite{supercool}).  The results show that, for any given  temperature there will always
exist transition regions where the local temperature is close to the critical temperature. In these regions one would expect thermal effects to play an important role. They may, in fact, dictate the transition from multi-fluid to single-fluid dynamics.

Perhaps not surprisingly---although the idea may seem somewhat unorthodox---we can use the entrainment to account for the presence of thermal excitations. As the idea takes a lead from the two-fluid model for Helium, it is instructive to first consider this case. The dynamics of superfluid Helium is easy to
understand if one starts from a system at zero temperature. Then
the dynamics must be entirely due to the quantum condensate; we have a single
quantum wavefunction and the momentum of the flow follows directly from the
gradient of its phase. This immediately implies that the flow is irrotational (we will address this later).
At finite temperatures, we must account for thermal excitations (e.g. phonons). Effectively, not all 
atoms remain in the ground state. A second dynamical degree of
freedom arises since the excitation gas may drift relative to the condensate. This motivates the two-fluid model---dating back to classic work
of London and Tisza (see \cite{wilks} for a nice discussion)---based on a distinction between a ``normal''
fluid component and a superfluid part \cite{khalatnikov,wilks,putterman}. 

The model we now advocate identifies the atoms (n) and treat the excitations (s) as a massless entropy component \cite{prix,sfhel}. This is different from the traditional approach from, for example, \cite{khalatnikov,wilks,putterman}. but it is easy to relate the two pictures. First of all, we identify the drift velocity of the quasiparticle
excitations in the two models. This is the variable that introduces
the ``two-fluid'' dynamics. Also, since it represents the part of the flow that is
affected by viscosity this components has a clear physical interpretation. In the standard
model this velocity, $v_\N^i$, is taken to be associated with the ``normal
fluid'' component. In the alternative framework, the excitations are associated
with the entropy of the system, which flows with $v_\s^i$. The two
quantities should be the same, so we identify
\begin{equation}
v_\N^i = v_\s^i \ .
\end{equation}

The second fluid component---the ``superfluid''---is commonly  associated with a
``velocity'' $v_\S^i$,  directly linked to the gradient of
the phase of the superfluid condensate wave function. This is,
in fact, a rescaled momentum and (as discussed in  \cite{prix}) we should identify
\begin{equation}
v_\S^i = \frac{p_\n^i}{m} \ .
\end{equation}
where $m$ is the atomic mass.
Combining \eqref{mom_def} with these identifications, we get
\begin{equation}
\rho v_\S^i = \rho \left[ \left(1 - \varepsilon\right) v_\n^i + \varepsilon
              v_\N^i \right] \ ,
\end{equation}
where $\varepsilon = 2\alpha/\rho$, with $\rho$ the total mass density. The total mass current is then given by
\begin{equation}
\rho v_\n^i = \frac{\rho}{1 - \varepsilon} v_\S^i -
\frac{\varepsilon \rho}{1 - \varepsilon} v_\N^i \ ,
\end{equation}
and
if we introduce the superfluid and normal fluid densities;
\begin{equation}
\rho_\S = \frac{\rho}{1 - \varepsilon} \ , \qquad \mbox{ and } \qquad
\rho_\N =  - \frac{\varepsilon \rho}{1 - \varepsilon} \ ,
\end{equation}
we arrive at the text-book result:
\begin{equation}
\rho v_\n^i = \rho_\S v_\S^i  + \rho_\N v_\N^i \ .
\label{mflux}
\end{equation}
Obviously, it is also the case that $\rho = \rho_\S + \rho_\N$. This completes the
translation. 

The comparison  shows that the variational approach identifies natural physical
variables; the average drift velocity of the excitations and the total
momentum flux. Since the system can be ``weighed'' the mass density $\rho$
also has a direct interpretation. Moreover, the variational derivation
identifies the truly conserved fluxes, see \eqref{continue}.
In contrast, the orthodox model uses
quantities that only have statistical meaning \citep{landau59:_fluid_mech}. The density
$\rho_\N$ is inferred from the mean drift momentum of the excitations. There is no ``group'' of excitations that can be identified with this
density. Since the superfluid density $\rho_\S$ is inferred from $\rho_\S =
\rho-\rho_\N$, this is a statistical concept, as well.
Furthermore, the two velocities, $v_\N^i$ and $v_\S^i$, are not individually
associated with a conservation law. From a practical point of view,
this is not a problem. The various quantities can be calculated from
microscopic theory and the results are known to compare well to experiments.
At the end of the day, the two descriptions are (as far as applications are concerned)
identical and the preference of one over the other is very much a matter of convention.

The comparison between the two formulations highlights a new concept: the entrainment between entropy and matter, which represents the effective mass of the thermal component. 
Comparing to \eqref{mom_def} we see that we have
\begin{equation}
\alpha = - \frac{\rho_\N}{2} \left( 1 - \frac{\rho_\N}{\rho} \right)^{-1} \ .
\end{equation}
This suggests that the entrainment coefficient $\alpha$ diverges as the temperature
increases towards the superfluid transition, as $\rho_\N \to \rho$. This may seem peculiar (as Nature abhors singularities), but it is simply a
manifestation of the fact that the two fluids must lock together as one passes
through the phase transition. The key point is that the model remains non-singular as long as
$v_i^\n \to v_i^\s$ sufficiently fast as the critical temperature is
approached. 

Returning to the neutron star problem, we may now add a third prescription for the entrainment to our arsenal. This approach involves a straight translation of \eqref{mflux} and the introduction of the superfluid velocity as the scaled momentum. We then have
\begin{equation}
V^i_\n = {1\over m_\n} p^i_\n = {1\over m_\n} \nabla^i \phi
\label{sfvel}
\end{equation}
This has the advantage of being directly linked to the gradient of the phase of the  wavefunction of the quantum condensate, $\phi$. This  leads to the mass-flux for each component
\begin{equation}
\rho_\x v^i_\x = \rho_{\x\x} V^i_\x + \rho_{\x\y} V^i_\y
\end{equation}
introducing the (symmetric) mass density matrix, $\rho_{\x\y}$. This decription was used in seminal work on superfluid neutron star dynamics \cite{als,mendell1}. Comparing to \eqref{mom_def} it is easy to show that we must have
\begin{eqnarray}
\rho_\n &=& \rho_{\n\n} + \rho_{\n\p} \label{reln}\\
\rho_\p &=& \rho_{\p\p} + \rho_{\p\n}
\label{relp}
\end{eqnarray}
and it also follows that 
\begin{equation}
\alpha = - {1\over 2} {\rho_\n \rho_\p \rho_{\p\n} \over \rho_{\n\n} \rho_{\p\p} - \rho_{\p\n}^2}
\label{alphel}
\end{equation}

The discussion of thermal excitations  allows us to make progress on the discussion of the dynamics near the superfluid transition point. The problem has been considered by, in particular, Gusakov and collaborators  \cite{gus1,gus2,gus3,gusha1} (see also \cite{fint}). Intuitively, it is clear what we need to do: We have to add thermal excitations to the model.  The outer core of a neutron star would then require (at least) four  distinct components;  neutrons, protons, electrons and the
entropy (representing the thermal quasiparticle excitations). While it is straightforward to write down the equations for such a system, solving them would be messy so it makes sense to (at least as a start) ignore two of the
relative degrees of freedom. First, we insist on charge neutrality (as before) by locking the  protons and the electrons. Second, we ignore the relative heat flux by locking the quasiparticles to the ``normal'' component, represented by the electrons. This, again, leaves us with a two-fluid problem. To get an impression of how this works, consider the key contribution from Gusakov and Haensel \cite{gusha1}, who
expressed the finite temperature effects in terms of the mass density matrix. Labelling the thermal excitations by an index qp, the  total non-relativistic fluxes can then be written
\begin{eqnarray}
\rho_\n v_\n^i &=&
\rho_{\rm nn} \,   V^i_{\rm n}  +
\rho_{\rm np} \,  V^i_{\rm p} + (\rho_{\rm n} - \rho_{\rm nn}-
\rho_{\rm np}) \,  v^i_{\rm qp} \,, 
\label{jn}\\
\rho_\p v^i_{\rm p} &=& 
\rho_{\rm pp} \, V^i_{\rm p} +
\rho_{\rm pn} \,  V^i_{\rm n}+(\rho_{\rm p} - \rho_{\rm pp}-
\rho_{\rm pn}) \, v^i_{\rm qp} \ , 
\label{jp}\end{eqnarray}
(here, it worth noting that  $v^i_\mathrm{qp}$ is a true transport velocity, in contrast to $V_\x^i$). 
The key points are: (i) the components of the mass-density matrix are now temperature dependent, and (ii) the relations \eqref{reln} and \eqref{relp} only hold at $T=0$. As in the case of \eqref{alphel}, the model can be translated to make contact with the variational derivation \cite{ft}. We then identify a``thermal effective mass'' that diverges as we approach the critical temperature. The formal divergence may seem troubling but an explicit example demonstrates that the model leads to the expected behaviour.  Far below the critical temperature the system exhibits a second sound, but this degree of freedom is quenched as one approaches the edge of the superfluid region leaving only the familiar sound waves in the normal fluid region \cite{ft}.

The impact of finite temperature effects on neutron star dynamics as not yet received the attention the issue deserves. The notable exception is a set of papers on the r-mode instability window \cite{sfrT1,sfrT2,sfrT3}, which demonstrates the importance of further work on the problem.

\section{Vortex dynamics}

The fluid equations we have considered do not---at least not explicitly---bring out one of the defining features of a superfluid system. Because the superfluid ``velocity'' follows from the gradient of a scalar phase, it should be irrotational. The fluid equations do not reflect this. This is as it should be, because macroscopic superfluids mimic bulk rotation by forming a dense array of quantised vortices and as long as we average over a large enough region the equations we have written down should be appropriate. Of course, the presence of vortices adds features to the problem. 

At the fluid dynamics level,  we introduce the 
angular frequency in the usual way; 
\begin{equation}
2 \Omega_\x^i = \epsilon^{ijk} \nabla_j v_k^\x \ ,
\end{equation}
As a result, on the macroscopic level we identify (assuming that the length scale considered is
sufficiently small that we can treat $\varepsilon_\n$ as constant)
\begin{equation}
\epsilon^{ijk} \nabla_j p^\n_k = 2 m_\n [ \Omega_\n^i + \varepsilon_\n (\Omega_\p^i - \Omega_\n^i)] \ .
\label{mac}\end{equation}
Meanwhile, at a mesoscopic level (where vortices are resolved) the circulation of the neutron momentum is quantised. 
Representing the mesoscopic ``momentum'' 
 by $\bar{p}_\n^k$, we have 
\begin{equation}
{1 \over m_\n} \epsilon_{ijk}\nabla^{j} \bar{p}_\n^k = \kappa_i = { h \over 2 m_\n} \hat{\kappa}_i \ ,
\label{mes}\end{equation}
where
\begin{equation}
\kappa = h/2m_\n\approx 2\times 10^{-3}\ \mbox{cm}^2/\mbox{s}.
\end{equation}
and the vortices are aligned in the direction of $\hat \kappa^i$. In practice, we expect to obtain \eqref{mac} from \eqref{mes}. Denoting the involved averaging procedure
by angular brackets we then  have 
\begin{equation}
\langle \epsilon^{ijk} \nabla_j \bar{p}^\n_k \rangle = \mathcal N_\v m_\n \kappa^i = \epsilon^{ijk} \nabla_j p^\n_k
\label{average1}\end{equation}
and we see that the vortex density (per unit area) is given  by
\begin{equation}
\mathcal N_\v \kappa = 2  [ \Omega_\n + \varepsilon_\n (\Omega_\p - \Omega_\n)] \ .
\label{n_v}\end{equation}

Formally, the fluid equations represent a
macroscopic average over a large number of individual fluid elements.
This is a natural approach for neutron stars where the focus is on dynamics at length scales vastly larger than the inter-particle separation. Of course, in order to assign values to the various parameters (like the entrainment)
one must resolve the problem at the microscopic level. This then raises
issues concerning the averaging required to make contact with 
the macroscopic level. When we add issues related to the  vortex dynamics, the situation is made 
even more complex. We have an intermediate ``mesoscopic'' 
level, sufficiently large that one can average over a large collection of 
particles (and  discuss ``fluid'' dynamics in a sensible way) and yet small
enough that one can resolve the individual vortices. The macroscopic equations follow from an averaging over a large set of vortices, leading to the smooth 
equations of motion  (\ref{euler}). 
To get an idea of the scales involved, it is useful to relate the vortex density to a typical neutron star (bulk) rotation rate; 
\begin{equation}
\mathcal N_\v \approx 6\times 10^5 \left( {10~\mathrm{ms} \over P}\right)\ \mathrm{cm}^{-2}
\end{equation}
where $P$ is the observed rotation period. This suggests that that the kind of averaging we have in mind would lead to fluid elements at the centimeter scale. It is also clear that the model we have outlined assumes that the vortices form an an aligned array. This is a debatable assumption, but it is a natural starting point.

\subsection{Mutual friction}

The vortex dynamics, both collective and individual, impact on the large-scale behaviour of neutron star superfluids. From a conceptual point of view, we need to understand how different effects associated with the vortices enter the fluid equations and to what extent this affects observed phenomena. The so-called mutual friction is key to this discussion. The classic argument leading to the vortex mutual friction (due to seminal work by Hall and Vinen in the 1950s \cite{hallvinen})
involves the scattering of thermal excitation (phonons) off of the normal fluid vortex cores in superfluid Helium. Mathematically, the neutron star problem is solved using the same strategy, but the physics argument is very different \cite{als,mendell1,trevor}. 

At the mesoscopic level, the motion of a neutron vortex is affected by 
fluid flow past it (known as the Magnus ``force'', see \cite{trevor}). At the same time, 
the circulation associated with the vortex induces circular flow in the 
protons because of the entrainment. The typical scale for this induced flow is related to the coherence length of the superfluid condensate, or order 100~fm. In effect,  the superfluid exhibits ``fluid'' behaviour on a much finer scale than one would expect above the superfluid transition temperature (where the size of a fluid element is dictated by the mean-free paths of the particles involved). The upshot of this is that a normal fluid component cannot ``react'' to the vortex circulation on the mesoscopic scale.  
This is essentially what happens in the Helium problem. For neutron stars we have another option---indeed, expected from the physics, see figure~\ref{pair}---associated with the fact that the protons in the star's core will also form a condensate. This means that they can be dragged along with the neutron circulation, due to the entrainment, while the electrons can not (at least not as a collective). The induced proton flow
generates a magnetic field on
the vortex, and the scattering of electrons off of this field 
acts resistively on the vortex motion \cite{als}.
The main friction mechanism in the stars is (yet again) different as it involves excitations of Kelvin waves on the vortices due to interactions with the lattice nuclei \cite{kelv1,kelv2,vg}.

In order to implement the mutual friction in the fluid equations we first of all consider the associated resistive force to be proportional to the difference in velocity between the vortex (moving with $v_\v^i$) and  the charged fluid flow. That is, 
we have
\begin{equation}
f_{i}^\mathrm{e} = {\cal R} (v^{\p}_{i} - v^{\v}_{i}) \ .
\label{resist}\end{equation}
acting per unit length of the vortex.
Neglecting the inertia of the vortex core,
and balancing $f_i^\mathrm{e}$ to the Magnus ``force'' we arrive at the mutual friction  
acting on the neutrons (per unit length of vortex) \cite{hallvinen, mendell2, trevor}
\begin{equation}
    f^{\rm mf}_i = {\cal B}^\prime \rho_\n \epsilon_{i j k} \kappa^j 
                   w_{\n\p}^k + {\cal B} \rho_\n \epsilon_{i j k} 
                   \epsilon^{k l m} \hat{\kappa}^j \kappa_l w^{\n\p}_m 
                   \ , \label{fmf1} 
\end{equation}
The same force acts on the vortex and an equal and opposite force 
affects the protons. Notably, the macroscopic force does not explicitly involve the vortex velocity. As discussed in \cite{als,mendell2,trevor} {estimates for the (dimensionless) mutual friction coefficients lead to}
\begin{equation}
{\cal B} = { { \cal R} \over \rho_\n \kappa} \approx  
4 \times10^{-4} \left( {\delta m_\p^* \over m_\p} \right)^2
\left( {m_\p \over m_\p^*} \right)^{1/2} \left( { x_\p \over 0.05} \right)^{7/6}  \left( { \rho \over 10^{14} \mathrm{g/cm}^3} \right)^{1/6} \ ,
\end{equation}
{along with, as the system is in the weak drag regime,}
\begin{equation}
{\cal B}^\prime \approx {\cal B}^2\ , 
\end{equation}
where $x_\p = \rho_\p/\rho$ is the proton fraction. The direct dependence on the entrainment (via the effective proton mass) is clear. Without entrainment, the proposed mechanism would not lead to the mutual friction. The relative importance of different proposed mutual friction mechanisms is discussed in \cite{vg,mfnew}.

In terms of completing the dissipative fluid equations, we note that the Magnus effect is already accounted for in the left-hand side
of  (\ref{euler}), so we only need to add the  averaged
version of the force (\ref{fmf1}) to the right-hand side.
In the particular case of a straight vortex array, the required 
averaging is easy: We
simply need to multiply (\ref{fmf1}) by the local number density of vortices 
per unit area, $\mathcal N_\v$, and we are done.

\subsection{Pulsar glitches}

As a dynamical example of mutual friction in action, it is natural to turn to the pulsar glitch phenomenon. Mature neutron stars tend to be extremely stable rotators, but many young systems exhibit  timing noise and (more or less) regular glitches, where the observed spin rate suddenly increases (see \cite{glrev} for a review and \cite{espinoza11} for a collection of
glitch data). These sudden spin-up events are followed by a slow relaxation 
towards the original spin-down rate. 

The archetypal glitching neutron star is the Vela pulsar, which has (since the first  event in 1969) exhibited a sequence of similar size glitches. The general understanding is that these glitches are a manifestation of the  superfluid neutron star  interior~\cite{baym69}. The idea was first put forward by Anderson and Itoh \cite{ai75} who explained a glitch as a tug-of-war between the
tendency of the neutron superfluid to match the spindown rate of the rest of the star
(by expelling vortices) and the impediment due to the vortices being ``pinned'' to crust nuclei. The vortex pinning prevents the
neutron superfluid from spinning down, leading to the development of a spin lag with respect to the rest of the star (which is spun down electromagnetically). 
The increasing spin lag  increases the Magnus force exerted on the vortices until---at some threshold limit---the vortices break free and the excess angular momentum is transferred
to the crust. This transfer of angular momentum is observed as a spin-up of the crust component (to which the star's magnetic field is locked). 

The model makes sense at the ``cartoon'' level, but it is probably fair to say that we do not yet have a  calculable model (see \cite{glitch1,glitch2,glitch3,glitch4,glitch5,glitch6,glitch7} for progress in this direction). This is not to say that there has not been progress. The relevant microphysics, especially concerning the  interaction between  neutron vortices and  crust  nuclei \cite{pin1,pin2,pin3,pin4,pin5,pin6,pin7,pin8}, is better understood, leading to a clearer idea of the location of the superfluid reservoir associated with the events. Observations support the notion that the vortices are (mainly) pinned in the crust and hence the angular momentum available is that associated with the crust superfluid. 

By considering the accumulated reversal in the spin down associated with glitches---an argument that requires regular events of (roughly) the same size---we may  infer that we need a superfluid reservoir with moment of inertia 
$I_{\rm n}/I \sim 1\%$ where $I$ and $I_{\rm n}$ are, respectively, the moments of inertia of the entire star and the
superfluid component \cite{link99}. This agrees well with the estimated moment of inertia of the crust (which is dominated by the free neutrons in the inner crust, see figure~\ref{pair}) for realistic equations of state \cite{rp94}. However, if the effective neutron mass is large (as suggested by the results shown in figure~\ref{entrain}) this explanation runs into trouble \cite{notenough1,notenough2}. 

As this is an important point, let us sketch the argument. First of all, we introduce a ``body'' averaged two-component model, including entrainment coupling. We then have
\be
\left( I_\p - \varepsilon_\n I_\n\right) \dot \Omega_\p + \varepsilon_\n I_\n \dot \Omega_\n = -a \Omega_\p^3  -\mathcal{N}_\mathrm{pin} -\mathcal{N}_\mathrm{mf}
\label{Jpdot}\ee
 where  the first term on the right-hand side represents the standard torque due to a magnetic dipole (the coefficient $a$ depends on the moment of inertia, the magnetic field strength and orientation), and 
\be
\left(1 - \varepsilon_\n\right) I_\n \dot\Omega_\n + \varepsilon_\n I_\n \dot \Omega_\p =  \mathcal{N}_\mathrm{pin} + \mathcal{N}_\mathrm{mf}
\label{Jndot}
\ee
We have added terms representing torques associated with
vortex pinning ($\mathcal N_\mathrm{pin}$) and dissipative mutual friction ($\mathcal N_\mathrm{mf}$). Noting that  the right-hand side of (\ref{Jndot}) vanishes for perfect pinning,
we see that---as long as the vortices remain pinned--- the crust  spins down according to
\be
\tilde I \dot \Omega_\p = - a \Omega_\p^3 \quad \mbox{where} \quad 
\tilde I = I_\p - {\varepsilon_\n \over 1 - \varepsilon_\n } I_\n 
\ee
The entrainment changes the effective moment of inertia and hence the  magnetic field inferred from the spin-down rate (neither of which are directly observed). 
Expressing the entrainment in terms of the (body averaged) effective neutron mass, we have
\be
\varepsilon_\n = 1 - { \left<  m_\n^* \right> \over m_\n}  \ \longrightarrow \ 
\tilde I = I -{m_\n \over  \left<  m_\n^* \right>}  I_\n 
\label{epsdef}\ee
If the effective mass is large, then the two components are essentially locked and the system spins down as one body ($\tilde I \to I$  in the limit where $\langle m_\n^* \rangle  \gg m_\n$). 
Turning to the rotation of the superfluid, we note that  $\Omega_\n$  changes even when vortices are pinned. From \eqref{Jndot} we have 
\be
\dot\Omega_\n = - {\varepsilon_\n \over 1 - \varepsilon_\n } \dot \Omega_\p = \left( 1 - {m_\n \over \langle m_\n^* \rangle} \right) \dot \Omega_\p
\ee
This impacts on the estimated glitch jumps because the spin-lag between the two components takes a longer time to develop
if the effective neutron mass is large. Working out the accumulated lag and assuming perfect angular momentum conservation during the glitch, we have (assuming $I \approx I_\p$)
\be
{\Delta \Omega_\p \over \Omega_\p} \approx {m_\n \over \langle m_\n^* \rangle }  \left({I_\n \over I}\right) { t_\mathrm{glitch} \over 2 \tau_c}
\ee
where $\tau_c$ is the characteristic age inferred from observations and $t_\mathrm{glitch}$ is the time between glitches. Introducing the activity parameter $A$ as in \cite{link99}---an argument that requires a system to exhibit regular glitches of roughly the same magnitude, as in the case of the Vela pulsar and the young x-ray pulsar J0537-6910 \cite{danai,ferd}---we have the constraint
\be
{I_\n \over I} \approx  2 \tau_c  \mathcal{A}  
 {\langle m_\n^* \rangle \over m_\n}  
\label{SFratio}
\ee
and we see that, if the (averaged) effective neutron mass is large,  the constraint on the moment of inertia inferred from  observations will be more severe. For a set of realistic equations of state and suggested pairing gaps, one finds values for $\langle m_\n^*\rangle /m_\n$  in the range $4-6$. This indicates that we cannot explain the observed glitches in terms of the crust superfluid \cite{notenough1,notenough2} (although see \cite{notenough3}). This region does not have sufficient moment of inertia to explain what we see. 

A plausible resolution to the problem is that some fraction of  the core superfluid is also involved in the glitch, just  enough to explain the observations. This could then allow us to use observed data to  constrain the singlet pairing gap for the neutrons (see figure~\ref{pair}), an interesting complement to the  constraints on the  superfluidity obtained from neutron star cooling.
Interesting, we may also turn the question around. Suppose the pairing gaps were firmly established. Then the observations would---at least in principle---allow us to infer the star's mass \cite{mass1} (see also \cite{glmass1,glmass2}). The superfluid would provide an internal weighing scale. This may seem somewhat far fetched, but it is a neat idea.

A precise understanding of glitch dynamics would involve both an explanation of statistical properties of a large set of events (an issue we have not considered here) and models able describe singular well resolved events (pulse-by-pulse, to the extent that this will ever be possible). The latter would constrain the mutual friction parameters, as this is the mechanism that couples the two components in the system once vortices unpin. Until recently, the best resolved event was the so-called Christmas glitch in the Vela pulsar \cite{vela1}, limiting the glitch rise time to less than a minute or so. Such an evolution would accord with the theoretical understanding of mutual friction. The recent results from \cite{vela2} is a potential game changer. Catching the Vela pulsar in the act of glitching, the new data suggest two  features that must be explained by theory: A slight dip in the spin rate preceeding the glitch and a short term over-shoot following it \cite{vela3,vg}. One possible explanation involves more than two components, but the discussion is only just beginning.

\subsection{Superfluid turbulence}

So far we have assumed that the vortices are (at least locally) straight and that they form an aligned array. Laboratory experiments suggest that both assumptions are dubious. Turbulent behaviour, associated with  vortex tangles, is common in superfluid systems and there is no reason why neutron stars should be  different \cite{packard,chevalier,turb}. 

\begin{figure}
\begin{center}
\includegraphics[width=0.7\columnwidth]{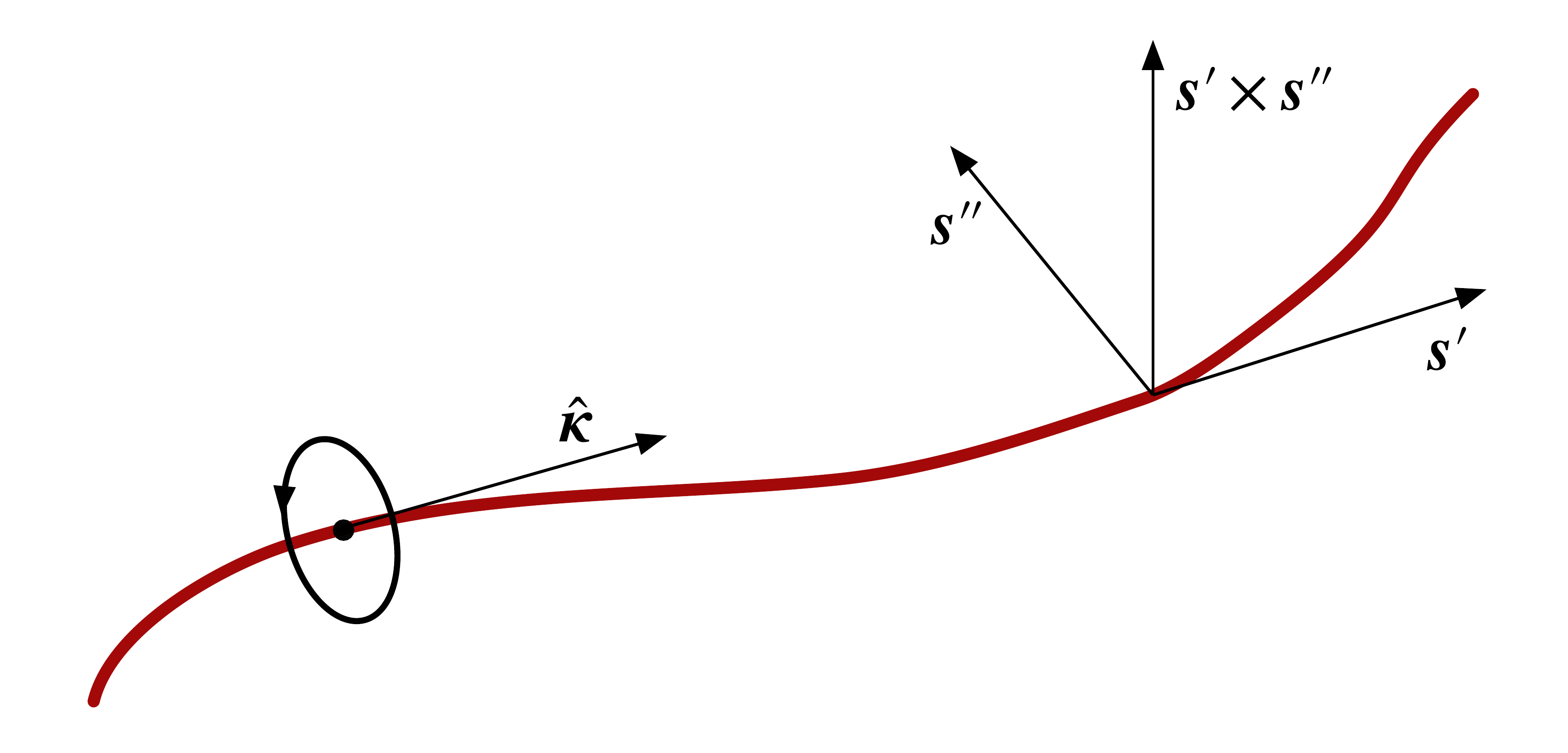}
\end{center}
\caption{A sketch of the vectors involved in the representation of the induced flow for a curved vortex.} 
\label{circ}
\end{figure}

If we want to model vortex tangles we, first of all, have to consider what happens when the vortices bend. The circulation of a curved vortex induces a flow that affects the motion of the vortex itself. The analysis proceeds as in the case of superfluid Helium, although we now have to account for  entrainment. As discussed in \cite{turb}, the vortex curvature induces a contribution to the 
(mesoscopic) neutron momentum
\be
{\bar{p}^\mathrm{ind}_i \over m_\n} = \nu \epsilon_{ijk}s^{\prime j} s^{\prime\prime k} 
\label{mesos}\ee
where
\be
\nu = { \kappa \over 4 \pi} \log \left( { b \over a_0} \right) 
\ee
A typical value of the size of the vortex core would be $a_0 \approx 100$~fm, while 
we can take $b$ to be given by the inter-vortex spacing.
We also have 
\begin{eqnarray}
s^\prime_i &=& \hat{\kappa}_i \\
s^{\prime\prime}_i &=& \hat{\kappa}^j \nabla_j \hat{\kappa}_i 
\end{eqnarray}
see figure~\ref{circ}.
Working out what this implies, we find that  the  difference between the two induced velocities is
\begin{equation}
v^{\n}_{i} - v^{\p}_{i} = \tilde{\nu}  \epsilon_{ijk} s'^{j} s''^{k}
\label{veldiff}
\end{equation}
where  
\be
\tilde{\nu} = \frac{1}{1- \varepsilon_{\n} - \varepsilon_{\p}} \nu
\ee
This then leads to  additional contributions---that should be added to (\ref{fmf1}))---to the mutual friction;
\be
f^{\rm ind}_i =  - \rho_\n \kappa \tilde{\nu}  \left[ {\cal B}^\prime \hat{\kappa}^j \nabla_j \hat{\kappa}_i 
+    {\cal B} \epsilon_{ijk} \hat{\kappa}^j \hat{\kappa}^l \nabla_l \hat{\kappa}^k \right]
\ee
effectively representing the tension of the vortices.

The introduction of the vortex curvature is important. From the Helium problem we know that 
there exists a critical relative 
flow above which oscillations are induced in the vortices rendering
the array 
unstable \citep{glaberson}, leading to a different form for the mutual friction \cite{gorter,hallvinen2} (see \cite{per1,per2} for interesting numerical simulations). 
This leads to the formation of a complex vortex tangle and a state 
of turbulence. 
In order to trigger the instability that drives turbulence, we need a flow along the vortices \cite{glaberson,trevor,trev2}.  This is easy to imagine in principle and fairly straightforward to arrange in the laboratory, but when would this happen in a neutron star? We need to induce a large scale flow that involves a component along the vortices. Two plausible scenarios  come to mind. The first involves free precession, where the precession motion induces circulation in the neutron star core. The results from  \cite{prec1,prec2} suggest that the vortex instability may prevent precession in systems where the vortices are strongly pinned. The second scenario relates to neutron star seismology, where the large scale oscillations may trigger local turbulence. It is also intriguing to entertain the possibility that the vortex array becoming unstable may explain the onset of glitches \cite{trigger1}. These are interesting ideas---see \cite{sfinst1,sfinst2,sfinst3,sfinst4} for further discussion---but many of the relevant details remain to be worked out.

\section{Oscillations and instabilities}

The discussion has taken us to the point where we are ready to discuss how superfluidity affects the problem of neutron star seismology, the general idea being that one would like to combine observations with theory models for  neutron oscillations to probe the physics of the star's interior. This is, inevitably, a challenging task, but as  there are  promising scenarios it is important to understand how superfluidity enters the discussion. There are different aspects to this. 

First of all, we may ask how the additional superfluid degree(s) of freedom impact on the oscillation spectrum. In general, 
different classes of oscillation modes can be---more or less clearly---assigned to different aspects of neutron star physics. Hence it is not surprising that  a superfluid
star has a distinct set of oscillation modes that arise because of the existence of the
second sound \citep{epstein,sfosc1,lee,cll,ac01,sfosc4,layers,lin,sfosc7,sfosc8,sfosc9,sfosc10} and it makes sense to ask if one may (at some point in the future) be
able to use observations, e.g. via gravitational waves \cite{glcprl}, to constrain the physics.

The second obvious aspect relates to the mutual friction. How does the vortex mediated friction impact on the mode oscillations? This turns out to be a crucial question in the context of the gravitational-wave driven (CFS) instability (see \cite{nareview} for a review). In fact, in the case of the fundamental mode of the star, the evidence is that the mutual friction may complete suppress the instability \cite{lindblom95,sfosc7}.
The situation is slightly different for the inertial r-modes, for which the situation may be more subtle. Still, there has been progress towards an
understanding of the nature of the r-modes in a superfluid neutron star \citep{lm00,sfr2,sfr3,sfr4,sfr5,pca}. The, increasingly detailed, effort to include finite temperature effects \cite{sfrT1,sfrT2,sfrT3} is notable in this respect.

A third relevant issue relates to the possibility that superfluidity may bring new features. Particularly interesting in this respect is the demonstration that superfluid systems may exhibit a two-stream instability, once the relative flow exceeds a critical level \cite{twost1,twost2,twost3,twost4,twost5}. The question of whether astrophysical systems are able to reach the critical level for the onset of this kind of instability remains open, but it is undoubtedly an interesting idea.

\subsection{Decoupling the degrees of freedom}

Considering the problem at the conceptual level, it is useful to  discuss the equations that govern a perturbed system (and which one would have to solve in order to determine the frequencies of the different oscillation modes). This also involves assumptions about the background one is perturbing with respect to. In the present case, as we want to highlight how the mutual friction enters the discussion, it makes sense to consider rotating stars but  (i) limit  ourselves to first order in the slow-rotation approximation (leaving out the centrifugal deformation which enters at second order), and (ii) focus on a background configuration where the two fluid rotate together. The general problem, with the components rotating at different rates already in the background is much more complicated (see for example \cite{sfdiff1}). 

Perturbing the equations of motion and working in a frame rotating with the angular frequency of the unperturbed star,  $\Omega^j$, we have
\be
\partial_t (\delta v^\x_i + \varepsilon_\x\delta w_i^{\y\x}) + \nabla_i (\delta \tilde{\mu}_\x + \delta \Phi)
+ 2 \epsilon_{ijk}\Omega^j \delta v_\x^k = \delta ( f^\x_i/\rho_\x)
\label{perteul}\ee
where $f^\x_i$ represents the mutual friction, and
\be
\partial_t \delta \rho_\x + \nabla_j (\rho_\x \delta v_\x^j) = 0
\ee
where $\delta$ represents Eulerian variations. For the usual two-component system, we have two fluid degrees of freedom and one might expect this to imply that the set of oscillation modes ``doubles'', with each mode of a normal fluid star having a new ``superfluid'' counterpart. Detailed work \cite{epstein,sfosc1,lee,cll,ac01,sfosc4,layers,lin,sfosc7,sfosc8,sfosc9,sfosc10} brings out this expectation (with some caveats that we will comment on later).

It is useful to ask what happens if we try to decouple the two degrees of freedom \cite{ac01}. The first natural degree of  freedom
represents the total mass flux.
Introducing the weighted sum
\be
\rho \delta v^j = \rho_\n \delta v_\n^j + \rho_\p \delta v_\p^j
\ee
and combining the two Euler equations accordingly, we find that
\be
\partial_t \delta v_i +  \nabla_i \delta \Phi + { 1 \over \rho} \nabla_i\delta p - { 1 \over \rho^2} \delta \rho \nabla_i p
+ 2 \epsilon_{ijk}\Omega^j \delta v^k = 0
\label{Eulav}\ee
where  $\rho = \rho_\n+\rho_\p$ and the pressure is obtained from
\be
\nabla_i p = \rho_\n \nabla_i \tilde{\mu}_\n + \rho_\p \nabla_i \tilde{\mu}_\p
\ee
In deriving \eqref{Eulav} we have used
\be
\rho_\n \nabla_i \delta\tilde{\mu}_\n + \rho_\p \nabla_i \delta \tilde{\mu}_\p = \nabla_i \delta p - \delta \rho \nabla_i \tilde{\mu}
=  \nabla_i \delta p - { 1 \over \rho} \delta \rho \nabla_i p
\label{chems}
\ee
where it has been assumed that the two fluids are in chemical equilibrium in the background;
$\tilde{\mu}_\n  =\tilde{\mu}_\p=\tilde{\mu}$.
Note also that, in the general case where the background fluids are not co-rotating, there
will be additional  contributions to \eqref{chems} associated with the differential rotation and the entrainment.

We also have the usual continuity equation
\be
\partial_t \delta \rho + \nabla_j (\rho \delta v^j) = 0
\label {conav}\ee
In effect, we  have two equations which are identical to the perturbation equations for a single fluid
body. In particular, we see that \eqref{Eulav} does not contain the mutual friction. This suggests that the co-moving degree of freedom is only damped by this mechanism by virtue of the coupling to the second degree of freedom.  

The other degree of freedom in naturally expressed in terms of the difference between the perturbed velocities: 
\be
\delta w^j = \delta v_\p^j - \delta v_\n^j
\ee
Combining  two Euler equations in the relevant way, we then have
\begin{equation}
(1-\bep) \partial_t \delta w_i + \nabla_i \delta \beta +
2 \bBp \epsilon_{ijk}\Omega^j \delta w^k - \bB \epsilon_{ijk}\hat{\Omega}^j \epsilon^{klm} \Omega_l \delta w_m = 0
\label{Euldi}\end{equation}
where we have defined
\be
\delta \beta = \delta \tilde{\mu}_\p - \delta \tilde{\mu}_\n
\ee
which represents the (local) deviation from chemical equilibrium induced by the perturbations. We have also introduced
the simplifying notation
\be
\bep = \varepsilon_\n/x_\p \ , \qquad \bBp = 1-\mathcal{B}'/x_\p \ , \qquad
\bB = \mathcal{B}/x_\p
\ee
where $x_\p=\rho_\p/\rho$ is the proton fraction.
Again,  equation \eqref{Euldi} does not appear to  couple the two degrees of freedom. The explicit  coupling is entirely due to the second continuity equation. It is 
natural to use the proton fraction to complement the total density $\rho$. If we do this, then we find that
\be
\partial_t \delta x_\p + { 1 \over \rho} \nabla_j \left[ x_\p (1-x_\p) \rho \delta w^j \right] +  \delta v^j \nabla_j x_\p = 0
\label{condi}\ee
This equation shows that the two dynamical degrees of freedom are coupled unless the proton fraction is constant \cite{sfosc4}.

In order to complete the formulation of the mode problem for superfluid neutron stars, we need to discuss two further issues; the equation of state and the boundary conditions. Starting with the equation of state, we need additional relations to close the system of equations. This typically introduces a less direct coupling between the fluid degrees of freedom.   Basically, we need to relate $[\delta p,\delta \beta]$  to $[\delta \rho, \delta x_\p]$ (or some other combination of these variables). For models where the two
fluids co-rotate in the background we have (in the general case there will be terms depending on the relative rotation and the entrainment here) 
\be
\delta \rho = \left( { \partial \rho \over \partial p} \right)_\beta \delta p +
\left( { \partial \rho \over \partial \beta} \right)_p \delta \beta
\ee
and
\be
\delta x_\p = \left( { \partial x_\p \over \partial p} \right)_\beta \delta p +
\left( { \partial x_\p \over \partial \beta} \right)_p \delta \beta
\ee

These relations give us an opportunity to discuss the impact of matter stratification of the star's oscillation spectrum. Focus first of all on the first relation and a single-fluid problem described by \eqref{Eulav}. This problem has two intuitive limits \cite{pantelis}. In the first, a displaced fluid element immediately equilibrates to the composition of its surroundings in the sense that the motion maintains (with $\Delta$ representing a Lagrangian variation and $\xi^j$ the associated displacement vector) 
\be 
\Delta \beta = \delta \beta + \xi^j \nabla_j \beta = 0 \quad \Longrightarrow \qquad \delta \beta = 0 
\ee
where the latter follows since the background is in equilibrium ($\beta = 0$). In this limit the equation of state is effectively a barotrope, and a non-rotating star has the usual f- and p-modes, with inertial modes like the r-mode becoming non-trivial due to the Coriolis force in a rotating star. Another ``simple'' limit is that of very slow nuclear reactions, where a moving fluid element does not have time to adjust before it moves on. In this limit we have $\Delta x_\p=0$. This introduces buoyancy and leads to the presence of g-modes associated with composition gradients \cite{pantelis}.  These arguments are standard. Now, superfluidity quenches the nuclear reactions so one might expect that the second argument would apply, but the answer turns out to be different as the variable $\delta \beta$ is now dynamical. The logic that leads to the buoyancy breaks down and the (usual) g-modes disappear \cite{lee,ac01}. Intuitively, the two fluids move relative to one another so a varying proton fraction does not lead to stratification (in the usual sense). We may, however, reintroduce a set of composition g-modes by considering the conditions beyond the density at which the muons first appear. We then have to consider a conglomerate of protons-electrons-muons, repesenting one of the two fluids, which will exhibit  buoyancy due to the varying  muon fraction \cite{muon1,muon2}.

The boundary conditions also involve elements of choice, reflected by the discussions in the literature (see \cite{cll,layers,lin}). The surface of the star is always identified by the vanishing of the Lagrangian variation of the pressure, $\Delta p=0$. This would also be the appropriate surface condition for a star with a superfluid interior since the superfluid is only present above the density of neutron drip.  However, there are situations where one may conceivably have to consider a superfluid surface, say for quark stars. In that case, it may be sensible to assume that the two perturbed fluids move together (in the radial direction) at the surface. 
For neutron stars, we need to focus on the junction conditions at the edges of the superfluid region. As we have already discussed in the context of finite temperature excitations, the dynamics in such regions may be characterised by the two fluids locking together \cite{ft}, so a natural option would be to introduce junction relations that facilitate this behaviour. There may, of course, be different physics arguments to consider. For example, at neutron drip the free neutron density vanishes and this may automatically enforce the required behaviour. At the end of the day we should let the physics dictate what the mathematics has to do. 

An oscillation problem of immediate relevance to observations involves the possible crust dynamics induced by the mechanism that leads to magnetar flares. Quasiperiodic oscillations seen in the x-ray tails from such events  (see \cite{qpo} for a review and further context)  are naturally interpreted in terms of the torsional oscillations of the crust. This, in turns, involves (due to entrainment) the crust superfluid. Different aspects of this problem are considered in 
\cite{magnet1,magnet2,magnet3,magnet4,magnet5}. As in the case of crust-driven glitches, the potentially strong entrainment coupling has significant impact on the crust oscillations and hence the interpretation of the observations.

Another topical problem related to neutron star oscillations---with immediate relevance for gravitational-wave astronomy---involves the possible imprint of superfluidity on the dynamical tide in an inspiralling binary system. This problem has been explored in \cite{sftide1,sftide2}, with suggestive results, but further effort is required to make the models realistic.

\section{Final remarks}

We have provided an introduction to different aspects that come to the fore when we consider the dynamics of superfluid system, paying particular attention to the implications of the two-fluid model and the role of entrainment for different problem settings. We have outlined the impact of superfluidity---and the associated vortex dynamics---on problems of immediate relevant for astrophysics, like pulsar glitches and neutron star seismology, adding references to the relevant literature that should help the interested reader dig deeper. Further effort would, indeed, be required to reach a precise understanding of specific technical aspects. The intention here was to provide a starting point rather than the final destination. 

In order to reach the journey's end (whatever that may be) we have quite a lot of work to do. This involves aspects we have  touched upon, like the nuclear physics and the main microphysics parameters (e.g. pinning and entrainment) and the averaging involved in the fluid model (especially involving vortices and turbulence). 
There are also aspects we have not considered, like the transport of heat in a superfluid system (e.g. the link to superfluid phonons \cite{phonon} and how this manifests itself in a fluid model) and features associated with the anticipated superconductivity of the protons in a neutron star core \cite{scon1,scon2,scon3}.  
Finally, we need to be mindful of the fact that a truly quantitative model---making contact with realistic microphysics---must be formulated in the context of general relativity. We are, in many ways, far from this goal. The relevant multi-fluid framework exists \cite{LivRev} and a number of simplified scenarios have been considered, but the challenges of truly realistic model building remain formidable.

\section*{Acknowledgements}

\noindent
My understanding of neutron superfluidity, as it is, has been developed through discussions with many great colleagues. Some are apparent from the listed references, other perhaps not. They are, however, too many to list (and I don't want to risk leaving someone out by mistake). I am also grateful for STFC funding through grant number ST/R00045X/1.

\end{document}